\documentclass[prb,twocolumn,amssymb,amsmath,secnumarabic,bibnotes,aps]{revtex4}
\usepackage{graphicx}
\usepackage{epsfig}
\usepackage{subfigure}
\newcommand{\unit}[1]{\;\text{#1}\,}
\newcommand{\vct}[1]{\boldsymbol{\vec{\textbf{#1}}}}
\newcommand{\cm}{\unit{cm}}			
\newcommand{\degree}{^{\circ}}		
\newcommand{\kg}{\unit{kg}}			
\newcommand{\km}{\unit{km}}			
\newcommand{\m}{\unit{m}}			
\newcommand{\mpers}{\unit{m/s}}		
\newcommand{\mperssquared}{\unit{m/s$^2$}}	
\newcommand{\nm}{\unit{nm}}			

\newcommand{\ns}{\unit{ns}}		
\newcommand{\radpers}{\unit{rad/s}}		
\newcommand{\revpers}{\unit{rev/s}}		
\newcommand{\T}{\unit{T}}			
\newcommand{\y}{\unit{y}}			

\newcommand{\uvct}[1]{\boldsymbol{\hat{\textbf{#1}}}}	
\newcommand{\gvct}[1]{\boldsymbol{\vec{#1}}}	
\newcommand{\guvct}[1]{\boldsymbol{\hat{#1}}}	
\newcommand{\cross}{\pmb{\times}}			
\def\centerbmp#1#2#3{\vskip#2\relax\centerline{\hbox to#1{\special
  {bmp:#3 x=#1, y=#2}\hfil}}}

\def\centereps#1#2#3{\vskip#2\relax\centerline{\hbox to#1{\special
  {eps:#3 x=#1, y=#2}\hfil}}}

\begin{document}
\title{Two Examples of Circular Motion for Introductory Courses in Relativity}
\author{Stephanie Wortel}
\author{Shimon Malin}
\email{smalin@mail.colgate.edu} \affiliation{Department of Physics, Colgate University, Hamilton NY 13346}
\author{Mark D. Semon}
\email{msemon@bates.edu}
\affiliation{Department of Physics, Bates College, 44 Campus Ave,
Lewiston ME 04240}

\begin{abstract}The circular twin paradox and Thomas Precession are presented in a way that makes both accessible to students in introductory relativity courses.  Both are discussed by examining what happens during travel around a polygon and then in the limit as the polygon tends to a circle. Since relativistic predictions based on these examples can be verified in experiments with macroscopic objects (e.g atomic clocks and the gyroscopes in Gravity Probe B), they are particularly convincing to introductory students.
\end{abstract}
\maketitle
\section{I. Introduction}

Experimental confirmations of relativistic effects are especially important for students in introductory relativity courses.  Each provides a particular situation that makes the abstract more concrete, and gives students an actual physical framework in which to understand what is (and is not) happening.

One of the most frequently discussed relativistic effects is the standard twin paradox.  In its most common form Twin A stays on Earth while Twin B travels at a constant velocity $\mathbf{v}$ to a neighboring star.  Twin B then moves into a new inertial frame and travels with a velocity $\mathbf{v}$ back to Earth.  The paradox is formulated by making two predictions which can't both be true.  On the one hand, since Twin B is moving in the frame of Twin A, Twin A should measure a clock held by Twin B to run slow. On the other hand, since Twin A is moving in the frame of Twin B, Twin B should measure a clock held by Twin A to run slow. The paradox, of course, is that when the two twins meet their clocks are face to face and each cannot have run slow relative to the other. 

The generally acknowledged resolution to the paradox is that the reference frames of Twin A and Twin B are not equivalent.  Twin B had to move from one inertial frame to another while Twin A didn't.  Consequently, since an accelerometer can distinguish between the two twins, there is no conflict with the Postulate of Special Relativity if the clock carried by Twin B runs slow relative to the clock carried by Twin A and not vice versa.

Following their presentation of the twin paradox many textbooks discuss its experiment confirmation in cyclotrons (e.g. Bailey et. al.,\cite{Bailey} Hay, et.al.\cite{Hay}) or with atomic clocks flown in airplanes (e.g. Hafele and Keating,\cite{HK72} Alley, et. al.\cite{Alley}).  However, neither experimental arrangement involves clocks moving along linear trajectories.  Rather, they more closely resemble what is called the ``circular twin paradox,'' which is formulated as follows: Suppose Twin A and Twin B are each on different rings and that the rings are rotating in opposite directions about a common axis through their centers.  One ring can either be just above the other or right next to it.  Assume each twin is holding a clock and that both clocks read zero when the twins first meet.  The paradox is the same as in the linear case: since Twin A is moving relative to Twin B, and Twin B is moving relative to Twin A, each should measure the other's clock as running slow compared to their own.  However, what distinguishes the circular twin paradox from the standard twin paradox is that now both twins accelerate in the same manner.  It is true that one accelerates in the clockwise and the other in the counterclockwise direction but, since time dilation depends only on the magnitude of the velocity, this can't matter. Consequently, unlike what happens in the standard twin paradox, the magnitude of acceleration measured on an accelerometer can't distinguish between the two travelers.  However, just as in the standard twin paradox, when the clocks meet and are face to face each cannot have run slow relative to the other.  There can be only one answer; what will that answer be?

Lightman, Press, Price and Teukolsky \cite{LPPT75} gave a short, formal solution to the circular twin paradox in 1975 and, in 2004, Cranor, Heider and Price \cite{CHP00} considered the paradox in more detail.  One purpose of Lightman, et. al. was to ``present an analysis that should be both mathematically and physically intelligible to beginning relativity students."\cite{CHP01}  Their approach, however, involved a detailed analysis of four different ways in which clocks on a rotating ring could be synchronized and a discussion of the consequences of each synchronization method.  The analysis became quite complex, but the authors said the complexity was necessary because, in their opinion, ``Special relativistic time dilation must be considered in the context of clock synchronization ..." and ``Time dilation will only be observed from a reference frame in which the clocks are appropriately synchronized ...". \cite{CHP01}

In this paper we show that the circular twin paradox can be resolved more simply by using the results of the linear twin paradox and that, contrary to what is claimed by Cranor et. al.,\cite{CHP00} it is not necessary to examine clock synchronization in the reference frame of either twin.  Our approach is to first consider what happens when the twins travel on concentric polygons and then to take the limit as the polygons tend to a circle.

The polygon method also can be used to discuss Thomas Precession \cite{Thomas} which, as far as we know, has never been included in an introductory relativity class.  Using elementary arguments involving rotations and boosts, we show that the coordinate axes of an inertial frame traveling around a polygon are rotated when they return to their starting point by an angle of $2\pi(\gamma -1)$ relative to their initial orientation.  Surprisingly, the direction of the rotation depends upon whether the polygon is traversed in the clockwise or counterclockwise direction.  In fact, Thomas Precession is one of the few effects in special relativity that depends upon the direction of the velocity rather than only upon its magnitude.  (Other examples occur in the addition of non-collinear velocities. \cite{Taylor}\, \cite{Costella}).  As noted by Uhlenbeck, ``even the cognoscenti of the relativity theory (Einstein included!) were quite surprised."\cite{Uhlenbeck}

The first experimental confirmation of Thomas Precession occurred just after the introduction of spin.  In fact, Thomas Precession played an important role in establishing the concept of spin (and the validity of quantum theory) since without it quantum mechanics is unable to correctly predict the fine-structure in the spectrum of hydrogen, hydrogen-like atoms, alkali atoms, etc.  

Perhaps less well-known is that another experiment to measure Thomas Precession is currently underway.  The Stanford-NASA satellite Gravity Probe B (GP-B), launched in 2004, contains a gyroscope predicted to precess, in part, due to Thomas Precession.  Consequently Thomas Precession is especially interesting to students because it involves the confirmation of a prediction of special relativity with a macroscopic object, and because the results of the gyroscope experiment in GP-B should be announced during 2007 and so are part of a contemporaneous physics research experiment that introductory students can understand.  Also, as we mention in our discussion, the technology involved in the gyroscope experiment is quite impressive and provides excellent topics for student projects and papers.

After discussing the major predictions of relativity -- time dilation, length contraction, etc. -- most texts discuss their experimental confirmation with muons, pions, and other elementary particles.  However, many students in introductory relativity courses are unfamiliar with these elementary particles and thus come to think of relativity as an interesting philosophical theory that has little to do with their everyday life.  Many are surprised to learn about relativistic effects on macroscopic objects such as atomic clocks and gyroscopes, and studying experimental confirmations made with these objects causes many students to take the concepts of relativity more seriously and to develop a new understanding of their importance.  Consequently, not only are the circular twin paradox and Thomas Precession interesting from a conceptual point of view, they also are especially effective at convincing students of the validity of relativity theory and the importance of its predictions.

The outline of this paper is as follows: In Section II we review the standard twin paradox and another paradox based on it. In Section III we discuss the circular twin paradox, its relation to the standard twin paradox, and its resolution.  In Section IV we discuss predictions based on the circular twin paradox and their confirmation in experiments with elementary particles in cyclotrons and macroscopic atomic clocks flown in airplanes. In Section V we discuss how acceleration can be treated in the various twin paradoxes.  In Section VI we derive Thomas Precession and illustrate how it affects a gyroscope orbiting the Earth.  In Section VII we discuss the anticipated confirmation of Thomas Precession with gyroscopes on the Stanford-NASA satellite Gravity Probe B.  We then use the gyroscope analogy to discuss spin and the experimental confirmation of Thomas precession in the spectrum of hydrogen. In Section VIII we summarize our results and discuss where they fit into an introductory relativity course.

One final note: the main body of this paper is written so that it can be understood by students in relativity courses for nonmajors, such as those using the texts by Mermin \cite{Mermin} or Baierlein. \cite{Baierlein}  The more quantitative parts have been removed from each section and placed in one of the three appendices.  In this sense the paper has two tracks, one designed for a more conceptually based course and the other for courses in which students have worked with Lorentz transformations and the relativistic form of Newton's Second Law; i.e., for courses such as the modern physics part of a standard calculus-based introductory course or any of the higher level courses in the physics curriculum.  Of course nonmajors with a good background in math and physics will find the appendices easy reading, and both types of students can follow up on the references for more detail about any of the subjects discussed.

\section{II. Linear Twin Paradoxes}

\begin{figure}
\begin{center}
\subfigure[\; Beginning of Trip as seen in Earth Frame.]{\epsfig{file=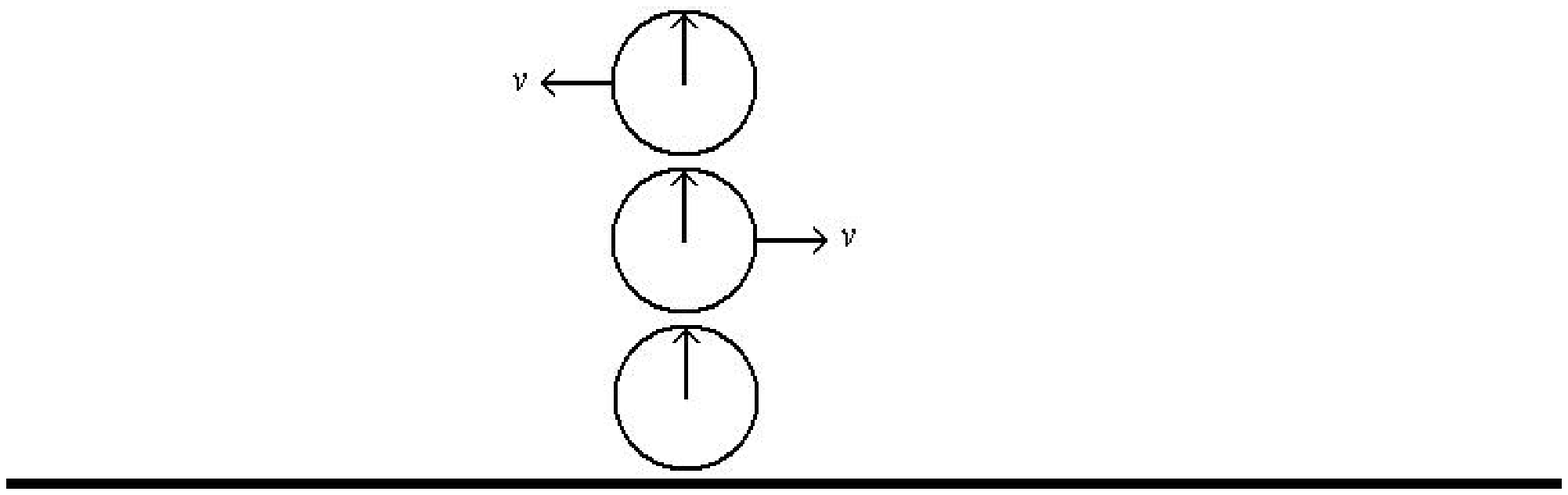, width=.35\textwidth}}
\subfigure[\; Turn Around as seen from Earth.]{\epsfig{file=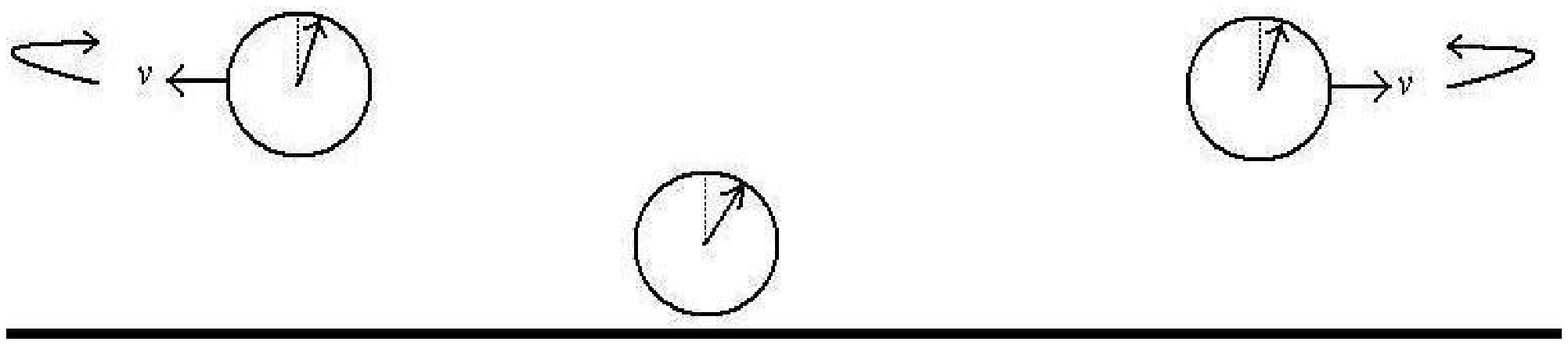, width=.3\textwidth}} \space
\subfigure[\; End of Trip as seen from Earth.]{\epsfig{file=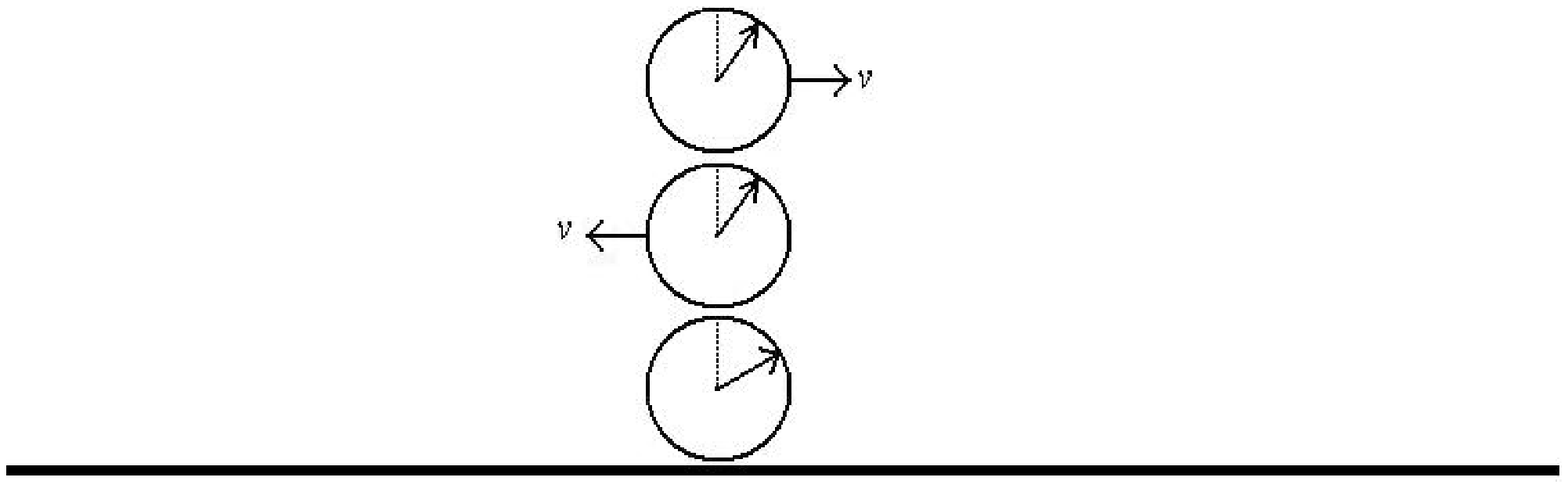, width=.3\textwidth}}
\end{center}
\caption{The ``Triplet Paradox."} \label{F1} 
\end{figure}

In order to understand the circular twin paradox it is helpful to first consider a slight generalization of the standard twin paradox to what can be called the ``triplet'' paradox.  Imagine triplets, one stationed on the Earth,\cite{ftnote3} one traveling with a velocity $\mathbf{v}$ to the right and the other traveling with a velocity $\mathbf{v}$ to the left. (See Figure \ref{F1}.) After each of the two traveling triplets has gone the same distance away from the Earth each moves into another inertial frame and travels back to the Earth with the same velocity $\mathbf{v}$.  Each triplet has a clock and the question is: When the two traveling triplets meet, how will their clocks compare?  Just as in the case of the standard twin paradox each traveler can argue that the other traveler's clock should run more slowly than their own.  However, in the triplet paradox both travelers undergo the same magnitude of acceleration so the reading on an accelerometer can no longer distinguish between them.  The direction of the accelerations can't matter since time dilation depends only upon the magnitude of the velocity and not upon its direction.  Therefore, the argument used to resolve the standard twin paradox won't resolve the triplet paradox.

Obviously the triplet paradox can be decomposed into two twin paradoxes.  Just as in the standard twin paradox, when each of the traveling triplets compares their clock with the clock on Earth each will agree that their clock is running slower than the Earth clock by the same factor of $\gamma$.  This means that each of the travelers will agree that the other traveler's clock is reading exactly the same time as their own and thus the paradox is resolved.  It is interesting to note that in this approach the resolution isn't obtained by directly comparing the clocks held by the two traveling triplets; rather, it comes from comparing each of the traveling clocks with a third clock, the one that remains on Earth.  (Stated more simply, in this approach the triplet paradox is resolved by proving the transitive relation: if $A = C$ and $B = C$ then $A = B$.  In our case $A$ and $B$ are the traveling clocks and $C$ is the clock that remains on Earth.)

\section{III. The Circular Twin Paradox}

\begin{figure}
\centereps{2in}{1.8in}{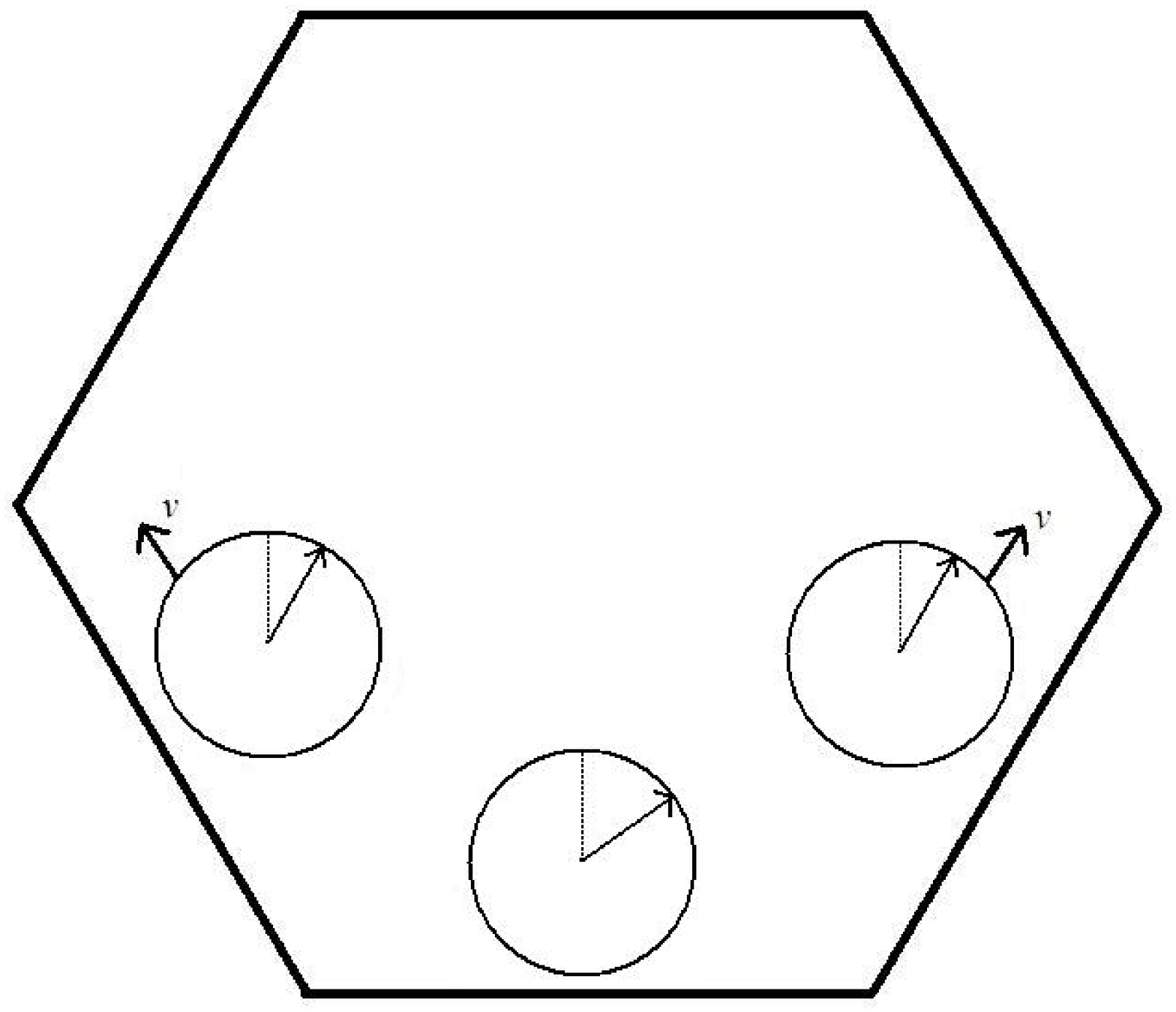}
\caption{Two clocks traveling away from an Earth based clock around an $N$-gon.} \label{F2} 
\end{figure}

The circular twin paradox can be resolved with the same method used in the triplet paradox. Consider a polygon with $N$ sides, as shown in Figure \ref{F2}.  Again consider the triplets, but now have each of the traveling triplets move with a speed $v$ along the polygon trajectory, one in the clockwise and the other in the counterclockwise direction.  Assume that when the traveling triplets first pass the Earth-based triplet all three clocks are compared and all read zero.  Along the side of the polygon at the bottom of Figure \ref{F2} the clock carried by each of the traveling triplets runs slow relative to the Earth clock by a factor of $\gamma$.  As the clocks move along each successive side of the polygon they continue to run slow by a factor of $\gamma$, so when the travelers are again across from the Earth triplet each of their clocks will be running slow by the same factor of $\gamma$.  Consequently, when they are again across from the Earth clock, the clocks held by each of the traveling triplets must read exactly the same.  All that remains is to take the limit as $N$ tends to infinity. When this is done the polygon tends to a circle and all of the results obtained for the polygon path are true for the circular path.  Note that this result is obtained by direct comparison of the traveler's clocks with the clock held by the Earth observer so there is no need to introduce a system of synchronized clocks in any of the reference frames involved.

It is interesting to note that from the point of view of a traveling triplet, the polygon path they follow is different from the polygon path measured by the Earth triplet.  If the Earth triplet measures each side of the polygon to be of length $L_0/N$, the traveler will measure each side to be of length $L_0/(N\gamma)$.  Consequently the traveling triplet will measure the total distance around the polygon to be less than that measured by the Earth observer.  Furthermore, if the Earth observer says that in limit as $N\rightarrow \infty$ the polygon tends to a circle of circumference $L_0=2\pi R$, then the traveling triplet will say that in this limit the polygon tends to a circle of circumference $L_0/\gamma=2\pi R/\gamma$. However, because lengths perpendicular to the motion don't contract, both triplets will agree that the radius of the circle is $R$.  Consequently, since the ratio of the circumference of their circle to its radius is less than $2\pi$, the traveler will conclude that the geometry of their frame is non-Euclidean. \cite{ftnote6} Einstein used an argument similar to this to prove that accelerating frames will have non-Euclidean geometries, and thus, that a theory of relativity generalized to include accelerated frames (i.e. general relativity) will have to be constructed using non-Euclidean geometry. \cite{Einstein} Thus the circular twin problem also can be used to show students an important aspect of general relativity and to give them an idea of what is meant by curved (or ``warped") space.

\section{IV. Experimental Confirmations of the Circular Twin Paradox}

The first experimental confirmation of the time dilation predicted in the circular twin paradox was reported by Hay et. al. \cite{Hay} in 1960. They put a Co$^{57}$ ``clock" on the surface of a cylinder whose radius was $0.4\cm$, and a Fe$^{57}$ receiver diametrically opposed to it at a radius $6.64\cm$.  When they rotated the unit at various angular speeds up to $500\revpers$ (corresponding to a linear speed of $7\cdot 10^{-7}$ c) they found the predicted and measured time dilations agreed to within an experimental uncertainty of two percent. \cite{Zhang}

In 1979 Bailey et. al. \cite{Bailey} performed a more accurate measurement using positive and negative muons traveling in circular orbits at the Muon Storage Ring at CERN.  The speed of these muons was approximately $0.9994$c  $(\gamma\approx 29.3)$.  Bailey et. al. measured a time dilation of the mean half lives that agreed with what was predicted by special relativity to within an experimental uncertainty of $0.1\%$.  Note that the experiments of Hay et. al. and Bailey et. al. involve exactly the same configuration as the circular twin paradox, which makes the paradox especially appropriate for discussing these two confirmations.

One of the advantages of including the circular twin paradox in an introductory relativity course is that predictions based on it have been confirmed with atomic clocks flown in airplanes.  Students who are only just learning about elementary particles are more impressed and convinced by experiments with macroscopic objects. In 1971 Hafele and Keating \cite{HK72} synchronized four cesium atomic clocks with a reference clock at the United States Naval Observatory in Washington, D.C. and then flew them around the Earth in commercial jets.  The clocks were flown first from east to west and then from west to east. (Four clocks were used on each flight in order to measure the average time elapsed and the experimental uncertainty of that average.)  When the clocks returned to the USNO they were compared with the reference clock.  The time difference predicted for the westward flight was $275 \pm 21\ns$ (i.e. $275\ns \pm 8\%$) while the observed time difference (averaged over the four clocks) was $273\pm 7\ns$ (i.e. $273\ns\pm 3\%$).  The overall experimental uncertainty is calculated by considering the standard deviations in each of the two results, and is usually expressed by saying that the observed and measured values agree to within ten percent. \cite{Rindler1}

Although the experiment of Hafele and Keating sounds simple enough when presented in this way there are several complications which may be worth discussing in class.  First, the predicted and experimentally measured discrepancies between the Earth-based and flying clocks actually arise from two effects: time dilation in special relativity, and a time dilation from general relativity which predicts that clocks at different heights in a gravitational field will run at different rates. Of course this doesn't alter the experimental confirmation of the time dilation predicted by special relativity; in fact, the agreement between the predicted and observed net discrepancies confirms the predictions of \textit{both} special and general relativity.  

Second, because commercial airplanes were used, the ground speed, latitude, longitude and altitude were not constant during each flight.  The westward flight was divided into $108$ intervals and the values of these four variables were recorded at the appropriate times.  The predicted time discrepancy was then obtained by numerically integrating over the trip's duration.  This is one of the reasons there is an uncertainty associated with the predicted result.

There is one additional detail that is sometimes worth discussing in class or assigning as a student project.  Since the Earth is rotating while the traveling clocks are in the air its rotational speed also must be taken into account.  Hafele and Keating present a clear and easy to follow discussion of this point in their papers. \cite{ftnote2} The result is that the relative velocity of the reference clock and the clock flown in the direction of the Earth's rotation will differ from that of the reference clock and the clock flown in the direction opposite to the Earth's rotation. In fact, the difference predicted for the eastward flight was $-40\pm 23\ns$ while the observed difference was $-59\pm10\ns$.  Although these numbers don't agree quite as well as the ones for the westward flight, they still confirm time dilation to within the experimental uncertainties.

It is interesting to note that the time difference between the Earth clock and the clocks flown eastward is negative while the difference between the Earth clock and the clocks flown westward is positive.  This means the clocks on the westward flight ran faster than the Earth clock, which means the experimenters on the westward flight aged more rapidly than the experimenters on Earth! At first this is somewhat surprising.  However, consider what is happening from the point of view of an inertial observer fixed in space on the axis of the Earth above the North Pole.  This observer will see the reference clock on Earth moving in a circular orbit eastward with a speed equal to the linear speed of a point on the surface of the Earth (at the location of the USNO).  They also will see the clock in the airplane moving in a circular orbit, but westward with a speed that is the resultant of the speed of the flying clock with respect to the Earth and the Earth's speed with respect to the inertial frame.  If the plane is traveling with a velocity relative to the inertial frame exactly equal but opposite to the Earth's linear velocity of rotation, then the inertial observer will see exactly the configuration of the circular twin paradox and conclude that both clocks should read the same time whenever they meet.  Reasoning in this way it is easy to understand what Hafele and Keating show explicitly: that if the clock flown in the westward direction is flying faster than the Earth's linear speed of rotation it will run slower than the clock fixed on Earth (i.e. the difference will be negative), whereas if the clock flown westward is flying slower than the linear speed of rotation of the Earth then the Earth clock will run slower than the clock in the airplane (i.e. the difference will be positive).  Because of the actual speeds of the jets used in the experiment, the clocks flown westward ran faster than the reference clock on Earth.
 
On the other hand, the clock flown eastward, in the direction of the Earth's rotation, will always be flying faster than the linear speed of rotation of the Earth and so it will always run slower than the reference clock on Earth.

\begin{figure}
\centerbmp{3.5in}{2.5in}{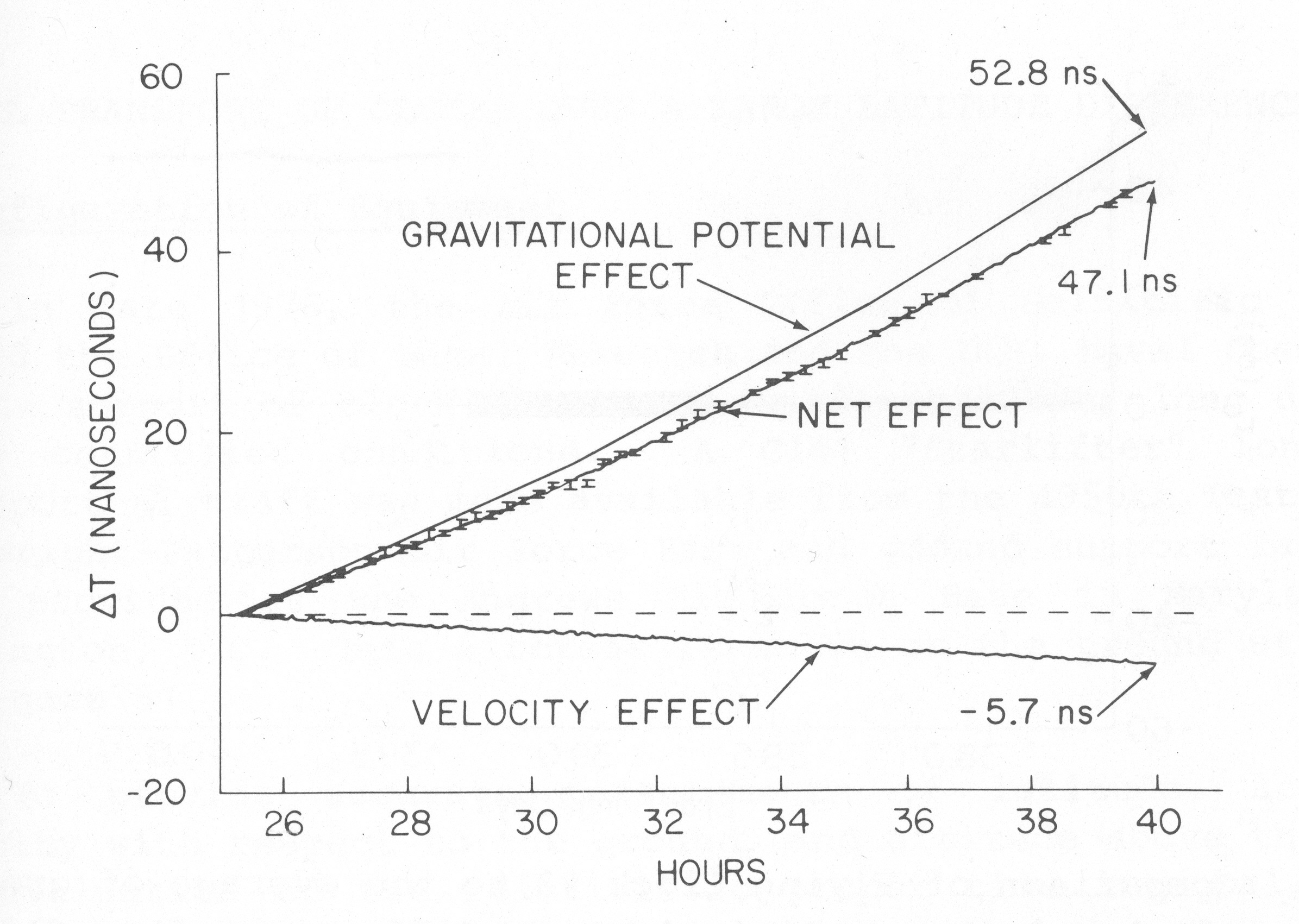}
\caption{Data from Alley et. al. \cite{ftnote22}} \label{A1}
\end{figure}

A more accurate experiment with atomic clocks was performed by C. O. Alley, et. al. in 1975.\cite{Alley}  His group put six atomic clocks, three cesium beam clocks and three rubidium gas cell clocks, on a U.S. Navy P3C anti-submarine patrol plane which made five fifteen hour flights in an elliptical (``racetrack") path over Chesapeake Bay.  Just as in the experiment of Hafele and Keating the ground speed, altitude, etc. of the plane were recorded, but this time continuously with both X-band and C-band radar so the integral of the time dilation could be calculated more accurately.  After landing, the plane was parked alongside a group of six identical reference clocks so a direct comparison could be made.  The clocks flown in the plane ran slower than the clocks that remained on the Earth.  The magnitude of the predicted difference was $47.1\pm 0.25\ns$ (i.e. $47.1\ns\pm 0.5\%$) and the magnitude of the measured difference was $46.5\pm0.75\ns$ (i.e. $46.5\ns\pm 1.6\%$).  A graph of their results, in which the prediction from special relativity (the ``velocity effect"), a prediction from general relativity (the ``gravitational potential effect'') and the net prediction are all plotted separately, is shown in Figure \ref{A1}.  The agreement between the observed and predicted discrepancies is quite remarkable, \cite{ftnote5} and confirms the predictions of \textit{both} special and general relativity.

\section{V. Acceleration in the Twin Paradoxes}

There still is the question of whether we are justified in ignoring acceleration in the twin, triplet and circular twin paradoxes.  In other words, does the rate of an ideal clock depend only upon its speed relative to an inertial frame, and is it independent of the clock's acceleration?  By ``ideal clock" we mean a clock based upon nuclear or particle decay rates, wavelengths or frequencies of atomic transitions, etc.

The experiments of Bailey et.al. \cite{Bailey} and Hay et.al. \cite{Hay} provide an excellent context in which to discuss this question.  In both cases ideal clocks were held in circular orbits by an applied magnetic field so their speed was constant.  This means that even though the clocks were accelerating $\gamma$ remained constant.  Therefore, if the time dilation predicted by special relativity agrees with the time dilation found experimentally, then the rate of an ideal clock has been shown experimentally to depend only upon its speed and to be independent of its acceleration.

As mentioned above, Hay, et.al \cite{Hay} found that when a Co$^{57}$ ``clock" was put on a rotating wheel the predicted and measured time dilations agreed to within an experimental uncertainty of two percent.  They also reported that their clocks experienced a constant acceleration of magnitude $10^4$ times the acceleration of gravity. \cite{ftnote20}  Consequently their experiment confirms that for accelerations of that magnitude, the rate of an ideal clock depends only upon its speed and is independent of its acceleration to within an experimental uncertainty of two percent.

When measuring the time dilation of the mean lifetime of muons in the Muon Storage Ring, Bailey et. al. \cite{Bailey} reported that the muons experienced accelerations of $10^{18}$ times the acceleration of gravity. (As we show in Appendix I, this value can be calculated from the basic parameters of the experiment.)  Consequently their experiments confirm that for accelerations of that magnitude, the rate of an ideal clock depends only upon its speed and is independent of its acceleration to within an experimental uncertainty of $0.1\%$.  Therefore experiments confirm that we are justified in ignoring the effect of acceleration on time dilation in the circular twin problem.

The question of acceleration in the other paradoxes can be handled in the same way.  If we assume that in the linear paradoxes the traveling twins are inside charged vehicles moving with a constant velocity away from the Earth, then they can pass from an inertial frame moving away from the Earth into an inertial frame returning to the Earth by being accelerated in a semi-circle by an applied magnetic field perpendicular to the plane of their path.  If they accelerate in this way, then the experiments discussed above confirm that the time dilation of these clocks depends only upon the constant value of $\gamma$ and is independent of their acceleration.  Similarly, when the twins are traveling around an $N$-gon, a constant magnetic field can accelerate their charged vehicle in a small circular arc from one side of the $N$-gon to the next.  The greater the magnetic field the smaller the arc.  The time over which the acceleration occurs also can be minimized by increasing the strength of the magnetic field.  As $N$ increases the change of direction $2\pi/N$ at each vertex decreases and the total change in direction ($2\pi$) is independent of $N$.

Using these results, we can summarize the various paradoxes from the point of view of the Earth observer, i.e., the observer who always remains in one inertial frame. In the standard twin paradox one twin moves away from the Earth and during this part of the trip the Earth observer measures the traveler's clock to run slow due to time dilation.  The traveling twin then accelerates into another reference frame heading back towards the Earth.  Experiments confirm that the acceleration necessary to change frames has no affect on the rate of the traveler's (ideal) clock. On the return trip the Earth observer again measures the traveler's clocks to run slow due to time dilation.  Consequently, when the traveler returns to Earth, the Earth observer predicts that they will be younger, and that their clock will have run more slowly by exactly the amount predicted by time dilation.  Obviously this analysis extends to the triplet paradox since it is constructed from two standard twin paradoxes, and to the circular twin paradox since what happens in this case is the result of what happens as the traveler moves from one straight line segment of the $N$-gon to the next which, in turn, is exactly what happens in the standard twin paradox.  Finally, as discussed above, the time dilation predicted in the circular twin paradox has been confirmed experimentally to a very high degree of accuracy.

Of course one can't help but wonder how the \textit{traveler} will explain why, when they return to their starting point, more time has elapsed on the Earth clock then on their own.  Because students often ask about this, and because including a description of the trip from the traveler's point of view completes the discussion of the circular twin paradox, we present it in Appendix II.  

\section{VI. Thomas Precession}

\begin{figure}
\begin{center}
\epsfig{file=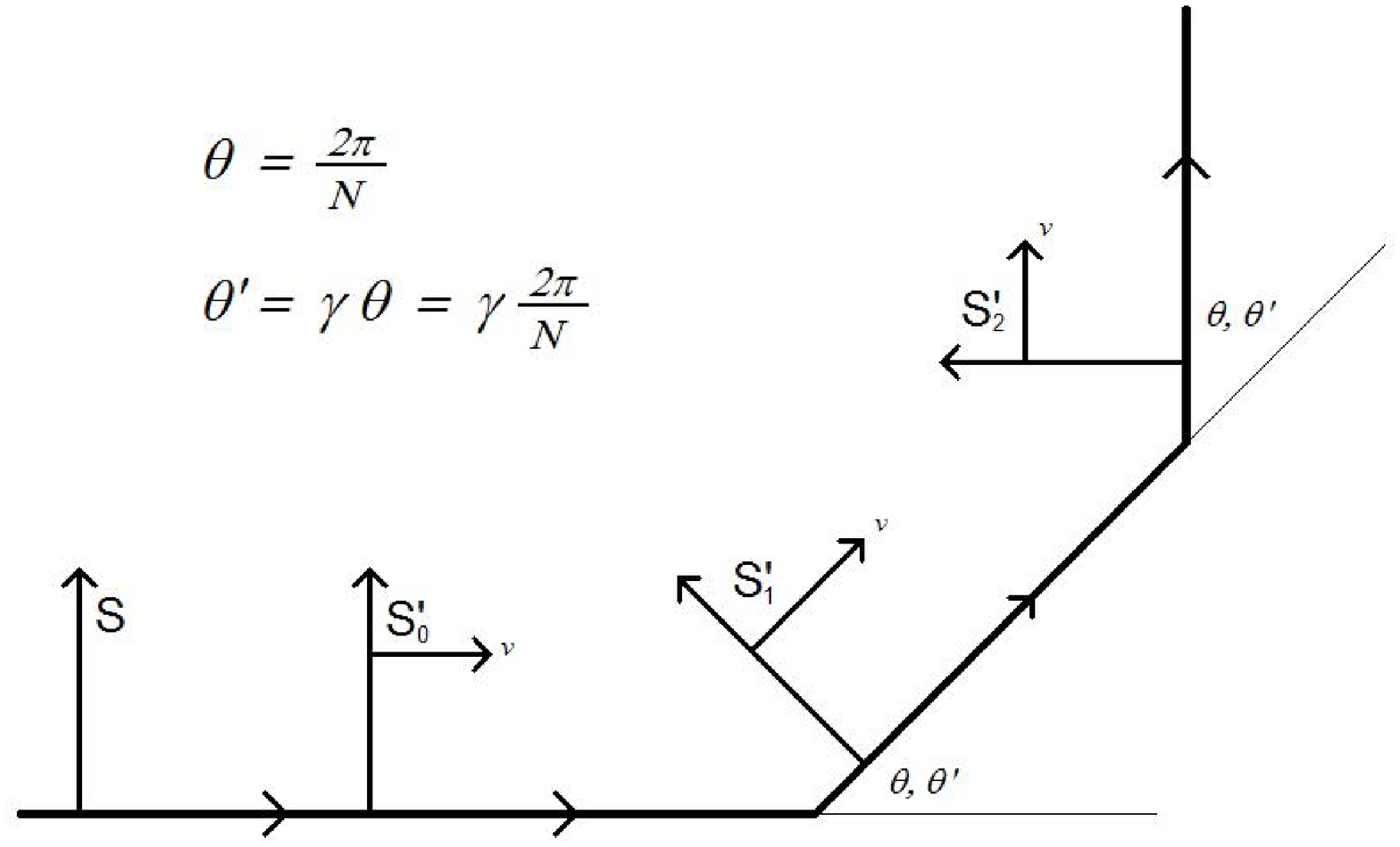, width=.4\textwidth}
\end{center}
\caption{Traveling the first two sides of the $N$-gon.} \label{F3} 
\end{figure}

We now show that if the traveling twin carries a gyroscope \cite{ftnote9} around an $N$-gon then, when they return to their starting point, they will observe the axis about which the gyroscope is spinning to have rotated, or precessed, through an angle of $2\pi(\gamma - 1)$. \cite{ftnote11}  In the limit as the $N$-gon becomes a circle, the traveler will observe the spin axis of the gyroscope to be precessing with an angular speed $\omega_T=2\pi(\gamma - 1)/P$, where $P$ is the time it takes the traveler to complete one trip around the $N$-gon.  The angle through which the spin axis moves relative to its initial direction is called the ``Thomas Rotation" angle and the rate $\omega_T$ of Thomas Rotation is called ``Thomas Precession."  It is important to remember that Thomas Rotation and Thomas Precession are effects observed by the traveling twin and not by the Earth twin, and that they are as real to the traveling twin as Coriolis and Centrifugal forces are to an observer in a rotating frame of reference.

To derive the Thomas Rotation and Precession, again consider a trip around the $N$-gon.  First, suppose the traveling twin walks around the $N$-gon with a constant nonrelativistic speed $v$ in the counterclockwise direction.  Also, suppose the traveler keeps re-aligning their $x'$-axis so that it points along the side of the $N$-gon on which they are traveling.  In order to do this, at each corner the traveler can first rotate their $x'$-axis through an angle $\theta$ given by $\tan\theta= 2\pi/N$ and then boost along the new direction.  We assume the boost doesn't change the speed of the traveler, only their direction. After one complete trip the traveler's $x'$-axis will have rotated through $2\pi$ and will point in exactly the same direction as when the trip began.

Figure \ref{F3} shows the trip in more detail.  The walking twin begins at rest with respect to the Earth in the frame $S$ and then boosts into the frame $S'_0$ traveling with a velocity $\vct{v}$ in their $x'$-direction.  As the traveler approaches the first corner they rotate their coordinate system and then boost into a new frame $S'_1$ moving along the second side of the $N$-gon.  As long as they are moving slowly, the angle between the horizontal axes of $S'_0$ and $S'_1$ is $\theta$ with $\tan\theta= 2\pi/N$.  In Figure \ref{F3}  we have approximated $\tan\theta$ by $\theta$ since our main interest is in what happens as $N\rightarrow\infty$ and in this limit the approximation $\tan\theta\approx\theta$ becomes exact.

Now suppose the traveling twin makes the same trip but at a constant relativistic speed $v$. Because of length contraction, they don't agree with the ground observer that they must rotate their $x'$-axis through $\theta = 2\pi/N$ in order to stay on the $N$-gon.  Rather, if the Earth twin measures $\tan\theta=y/x$ then, because the traveling twin observes a length contraction $x'=x/\gamma$ in the direction of travel, the angle through which they have to rotate is $\theta'$ where $\tan\theta'=y'/x'=y/(x/\gamma)=\gamma (y/x) = \gamma\tan\theta.$  This is shown in Figure \ref{F3} in which, as mentioned above, we have approximated  $\tan\theta'$ by $\theta'$.  Therefore, at each corner $\theta'=\gamma\theta=\gamma(2\pi/N)$ as $N$ gets large.

\begin{figure}
\begin{center}
\epsfig{file=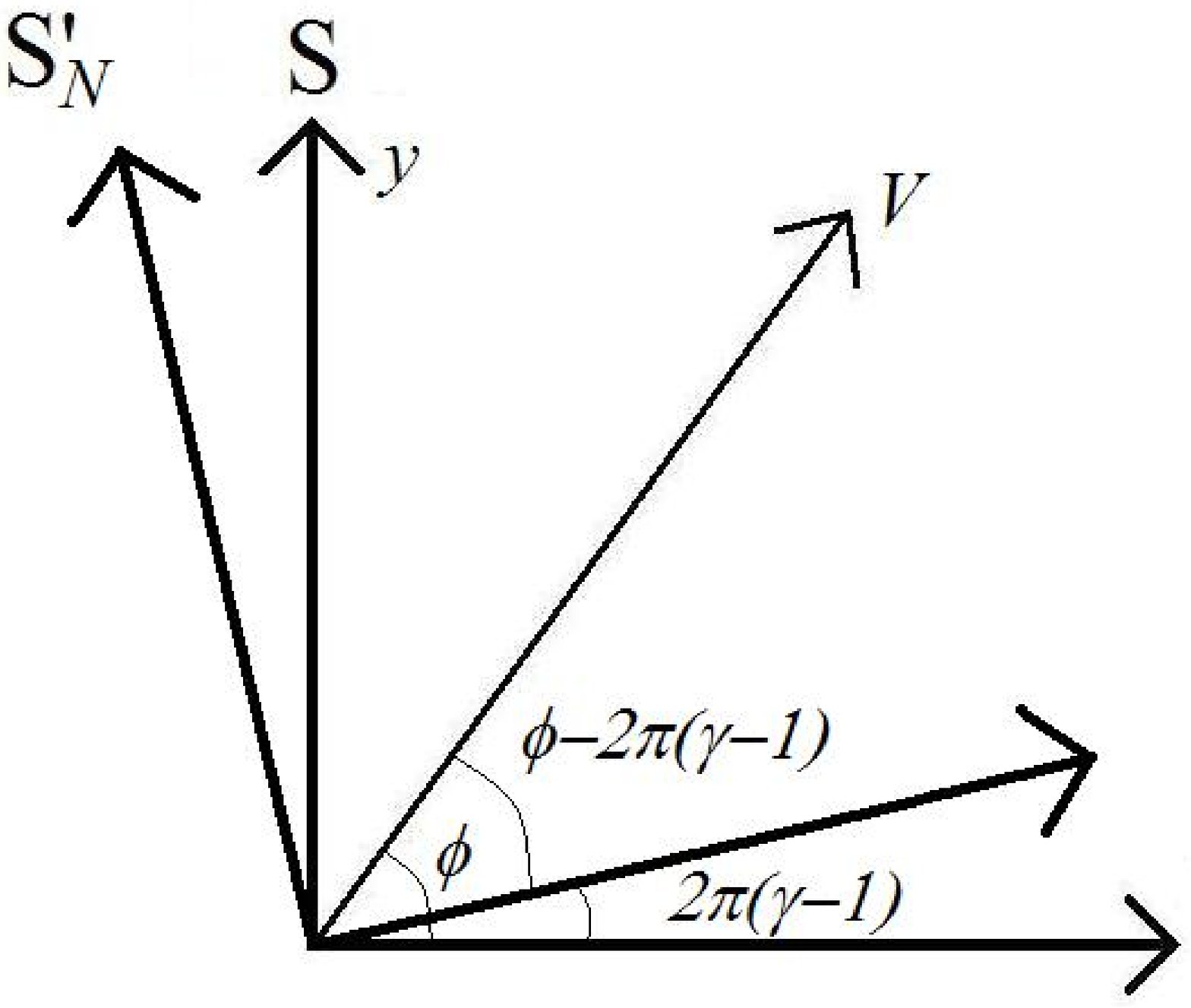, width=.4\textwidth}
\end{center}
\caption{The Thomas Rotation Angle $\phi- 2\pi(\gamma-1)$.} \label{F4} 
\end{figure}

When the traveling twin makes one (relativistic) round trip and returns to the point at which they began, their $x'$-axis will have rotated by an angle of $2\pi\gamma$.  Figure \ref{F4} shows the orientation of the final frame $S'_N$ with respect to the initial frame $S$ after one round trip. Consequently the Earth twin says the coordinate system of the traveling twin has rotated in the counterclockwise direction by an angle of $2\pi(\gamma-1)$. 

Suppose the traveling twin makes their round trip in a satellite carrying a gyroscope.  If the spin axis of the gyroscope $V$ originally made an angle of $\phi$ with the traveler's horizontal axis then, when the traveler returns to their starting point, $V$ will now make an angle $\phi- 2\pi(\gamma-1)$ with the traveler's horizontal axis.  This is shown in Figure \ref{F4}.  Since the traveler refers everything to their coordinate system, they will observe the spin axis of the gyroscope to have rotated in the clockwise direction by an angle $2\pi(\gamma-1)$. This is the Thomas Rotation.

In the limit as $N\rightarrow\infty$ the $N$-gon tends to a circle.  Since the Thomas Rotation is independent of $N$ we see that relative to the traveler, the gyroscope they are carrying rotates in the clockwise direction by $2\pi(\gamma-1)$ every time they make one trip around the circle in the counterclockwise direction.  If $P$ is the period of the orbit, then the traveling twin observes the gyroscope to precess clockwise in the plane of the orbit with an angular speed 
\begin{equation}
\omega_T=(2\pi/P)(\gamma-1)=\omega(\gamma-1), \label{TP0} 
\end{equation}
where $\omega=(2\pi/P)=(2\pi$ radians per cycle) $\times \;(1/P$ cycles per second) $=$ the angular speed (in radians per second) with which the twin moves around the circle.  This is the Thomas Precession.

If students are familiar with the vector description of circular motion (as presented, for example, in Reese \cite{Reese} or Boas \cite{Boas}) then the angular velocity $\gvct{\omega}_T$ of Thomas Precession can be written in a more general (and standard) form.  This is done in Appendix III.  Nonetheless, as we now show, students need only understand the basics of circular motion in order to derive a simple equation from which the important applications of Thomas Precession can be obtained.

If $s$ is the arclength traveled by the satellite when it moves through an angle $\theta$ in its orbit then $\theta=s/R$ when $\theta$ has the units of radians. Since $R$ is constant, the rate of change of $\theta$ is equal to the rate of change of $s$ divided by $R$.  Therefore, since $\theta$ changes by $2\pi$ every period,
\begin{equation}
\omega=\frac{2\pi}{P}=\frac{\Delta\theta}{\Delta t}=\Big(\frac{\Delta s}{R}\Big)\frac{1}{\Delta t}=\Big(\frac{\Delta s}{\Delta t}\Big)\frac{1}{R}=\frac{v}{R}.
\end{equation}

Putting this result into Eq. (\ref{TP0}) we have
\begin{align}
\omega_T&=\omega(\gamma-1)=\Big(\frac{\omega v^2}{v^2}\Big)\Big(\frac{R}{R}\Big)(\gamma-1)\\
&=\Big(\frac{\omega R}{v^2}\Big)\Big(\frac{v^2}{R}\Big)(\gamma-1)\\
\implies\omega_T &=\Big(\frac{va}{v^2}\Big)(\gamma-1)\label{TP'}
\end{align}
where $a=v^2/R$ is the usual centripetal acceleration.

The reason for writing Eq.(\ref{TP'}) in such a peculiar form is that in the nonrelativistic limit $\gamma\approx 1-v^2/2c^2$ and Eq. (\ref{TP'}) becomes
\begin{equation}
\omega_T =\frac{va}{2c^2}.\label{TP10}
\end{equation}

If a satellite is held in orbit by the gravitational force the acceleration of its reference frame is $a=\big(GM/R^2\big)$, where $R$ is the radius of the orbit and $M$ is the mass of the Earth.  In this case, since $a=v^2/R,\;v=\sqrt{GM/R}$.  If the satellite contains a gyroscope then, according to Eq. ({\ref{TP10}}), the traveling twin will see the gyroscope precess with an angular speed
\begin{equation}
\omega_T= \frac{1}{2c^2}\frac{GM}{R^2}\sqrt{\frac{GM}{R}}= \frac{1}{2c^2R}\Big(\frac{GM}{R}\Big)^{3/2}\negthickspace\negthickspace\negthickspace.\label{TP5}
\end{equation}

It's interesting to note that even though the Thomas Rotation angle depends upon $\gamma$, the Thomas Precession rate does not.  This is because $\omega_T$ is the quotient of the Thomas Rotation angle and the period, and the $\gamma$ factor in the latter cancels the $\gamma$ factor in the former.  Consequently, Thomas Precession depends only upon the radius of the orbit and is independent of the speed.

The same precession occurs in the reference frame of the electron in a hydrogen atom.  In this case the ``satellite" is the electron, the ``gyroscope" is the electron's spin vector and the acceleration from the Coulomb force replaces the gravitational acceleration.  As we discuss in the next section, when quantum mechanics is used to predict the spectrum of hydrogen certain calculations are easiest to perform in the electron's frame.  Once the desired result is obtained it is then transformed back into the proton's frame.  Originally this transformation was done without taking the Thomas Precession into account and it differed from the experimental result by a factor of two.  It was in order to account for this factor that Thomas did his original work. \cite{Thomas} He showed the problem was that what we now call Thomas Precession was not being included in the transformation from the electron's frame to the proton's frame.  Once the Thomas Precession of the electron's spin vector was taken into account the predicted wavelengths in the hydrogen spectrum agreed with those found experimentally.

Since the Coulomb force holds the electron in its orbit the acceleration in Eq. (\ref{TP10}) is $a=\big(ke^2/m_e r^2\big)$ where $k$ is a constant, $m_e$ is the mass of the electron and $e$ is the charge of the electron.  Replacing $GM$ in Eq. (\ref{TP10}) with $ke^2/m_e$  we find that in this case the magnitude of the Thomas Precession is
\begin{equation}
{\omega}_T= \frac{1}{2c^2r}\Big(\frac{ke^2}{m_e r}\Big)^{3/2}.
\end{equation}

\section{VII. Experimental Confirmation of Thomas Precession.}
\subsection{A. Gyroscopes Orbiting the Earth: Gravity Probe B}

During the period 1959-1960 G. Pugh \cite{Pugh} and L. I. Schiff \cite{Schiff} independently wrote papers pointing out that a gyroscope orbiting the Earth should precess.  They each found that three different effects, one from special relativity and two from general relativity, should contribute to that precession: the Thomas Effect, discussed above, the ``geodetic effect," which results from a general relativistic correction to Newton's theory of gravity, and the ``Lense - Thirring Effect" which results from the Earth's ``dragging" the local inertial frame in the direction of its rotation.  If the satellite is in a polar orbit, then the Lense - Thirring effect makes the gyroscope precess perpendicular to the plane of the orbit and in this way is separated from the Thomas and geodetic precessions, both of which are in the plane of the orbit. \cite{Holstein}  Consequently, just as in the previous experiments with atomic clocks flown in airplanes, a verification of the net precession in the satellite's orbital plane will confirm \textit{both} Thomas and geodetic precession.

Work on designing and implementing such an experiment began in 1962.  After the heroic efforts of many different people over many years, the satellite Gravity Probe B (GP-B) was launched in 2004. \cite{GPB} In principle, the experimental setup is quite simple.  Put a satellite containing a gyroscope and an optical telescope in orbit around the Earth.  Align both the gyroscope spin axis and the telescope to a guide star.  Keep the telescope pointed towards the guide star and measure the movement of the spin axis of the gyroscope for one year.  If the satellite is put into a polar orbit then the movement of the spin axis in the plane of the orbit should be that predicted by the sum of the Thomas and geodetic precession, and the movement perpendicular to the plane of the orbit should be that predicted by Lense and Thirring.

The immense practical obstacles to measuring these precessions become apparent when Eq.({\ref{TP5}) is used to calculate their magnitudes. The radius of the GP-B orbit is  $r=R_E+r_s=6.378\times10^6\m+6.493\times10^5\m=7.0274\times10^6\m$. \cite{radius}.  The mass of the Earth is $M=5.974\times10^{24}\kg$, the speed of light is $c=2.9979\times10^{8}\mpers$, and the gravitational constant is $G=6.672 \times 10^{-11}\unit{N$\cdot$m$^2$/kg$^2$}$.  Putting these numbers into Eq.({\ref{TP5}}) we find that $\omega_T=1.0145\times{10^{-12}}\radpers$.  The geodetic precession is twice this magnitude and in the opposite direction. \cite{Holstein} Consequently, the total precession in the plane of the orbit is predicted to be $3/2$ the magnitude of the Thomas precession, or $0.001834\degree$/year = $6.6$ arcseconds per year.  (It's gratifying to students that the previous equations result in the actual value of the net precession the experiment is designed to measure.)  The experimenters note that this number is approximately equal to the width of a human hair when viewed from $0.25$ mile away.  (Verifying this analogy makes a nice student exercise, as does constructing similar analogies with dimes, etc.)  The technology developed to preform this experiment is so remarkable that the precession is expected to be measured with an experimental uncertainty of better than $0.5$ milliarcseconds $=1.39\times10^{-7}$ degrees, or to within $0.01\%$. The results are expected to be announced near the end of 2007 so students who have studied Thomas Precession will have the background necessary to understand this contemporary (and newsworthy) verification of special relativity.
  
The history of Gravity Probe B and the technology developed to make such accurate measurements is a fascinating story, any part of which can easily be researched and presented as a student project. \cite{GPB} For example, the quartz gyroscopes used are the most spherical objects ever made, with radii constant to within $3\times10^{-7}$ of an inch. They are homogeneous to within $2$ ppm and coated with a layer of niobium that is $1270\nm$ thick.  The niobium is superconducting and the magnetic moment generated by the rotating quartz-niobium sphere is detected by SQUIDs which can measure magnetic fields on the order of $10^{-13}$ Gauss.  The gyroscopes are so free of their mountings that their spindown time is estimated at $15,000\y$.  The list of remarkable inventions and developments goes on, and any one of them can serve as the basis for a student report or project. 

\subsection{B. The Fine-Structure of Hydrogen}

The first confirmation of Thomas Precession occured in 1927 when it was used, along with the concept of spin, to derive the correct prediction of the spectrum of hydrogen. (A wonderful description of the historical context in which Thomas' papers appeared is given by Tomonaga. \cite{Tomonaga}) The way Thomas Precession entered the calculations is as follows: the potential energy of the electron associated with its spin is first calculated in the electron's rest frame.  It then has to be transformed back to the frame of the proton.  However, as we have seen, any result calculated in the electron's frame has to include the Thomas Precession factor when it is transformed back to the proton's frame.  Before this was done the predicted results didn't agree with what was found experimentally.  After the Thomas Precession was included they did, not only for the hydrogen atom, but also for hydrogen-like atoms, alkali atoms, etc.  As Tomonaga remarked, ``Thus all the fog of 1923-24 [was] completely cleared."\cite{Tomonaga}

\section{VIII. Conclusion.}

In this paper we have presented two examples which can be used in relativity courses for majors and non-majors.  Both involve circular motion, and both make predictions which can be confirmed in experiments with macroscopic objects.  The first, called the ``circular twin paradox," provides the theoretical basis for experiments confirming time dilation with atomic clocks flown in airplanes.  The second, ``Thomas Precession," is currently being tested with gyroscopes in the Stanford - NASA satellite Gravity Probe B (GP-B).  

Thomas Precession is rarely (if ever) discussed in introductory relativity courses.  We present a derivation based on length contraction alone, and then discuss the Thomas Precession predicted for gyroscopes on the GP-B satellite.  The gyroscope experiment makes Thomas Precession easy to visualize and provides a context in which to discuss the remarkable technologies developed in order to make the gyroscope experiment possible.  Since the results of this experiment are expected to be reported towards the end of 2007, discussing Thomas Precession makes a ``cutting edge" physics research experiment accessible to introductory students.

Both of these examples can be dicussed in a relativity course after time dilation and length contraction have been derived.  Their confirmation with macroscopic objects helps students accept some of the experimental and philosophical consequences of special relativity, and challenges their belief that relativity only affects elementary particles and is of no importance in everyday life or in our understanding of the more general concepts of time and space.

\section{Acknowledgments}

We thank Nathaniel Stambaugh for helpful discussions and for his expertise in making the figures.

\section{Appendix I: Calculating the Acceleration of the Muons in the Experiments at CERN.}

If students are familiar with the Lorentz transformations for the four-velocity and four-acceleration then calculating the acceleration in the experiments of Bailey et. al. \cite{Bailey} makes a simple but nice application that can be done in class or assigned as an exercise.

Bailey et. al. report that in their experiment the muons were moving in a circular orbit of radius $7.00\m$ with a constant speed of $v=0.99942$ c ($\gamma= 29.304$).  In this case the relativistic form of Newton's second law is 
\begin{equation}
\vct{F}= \frac{d(\gamma m\vct{v})}{dt} = \gamma m \vct{a}
\end{equation}
Consequently, since the force holding the muons in orbit is magnetic,
\begin{equation}
qvB=\gamma ma\implies a=\frac{qvB}{\gamma m}.\label{EA1}
\end{equation}
Using the value of $B = 1.472\T$ reported by Bailey et. al., and taking the mass of the muon as $206.7$ times the mass of the electron,  we find that $a= 1.28\cdot 10^{16}\mperssquared$.  However, the acceleration calculated in Eq.\,(\ref{EA1}) is the acceleration measured in the lab frame.  The acceleration experienced by the muons is the \textit{proper} acceleration which, for circular motion, is $\gamma^2\,a$. \cite{Rindler2}  Consequently, the muons experience an acceleration of $1.1\times 10^{19}\mperssquared$, or approximately $10^{18}$ times the acceleration of gravity.  This is the value stated by Bailey et. al. in their paper.

\section{Appendix II: How the traveling twin explains why their clock runs slow relative to the Earth Twin.}

In Section IV we described the twin, triplet and circular twin paradoxes from the point of view of the Earth observer.  Students often ask how the traveler describes the trip.  In this section we answer this question in the context of special relativity by elaborating on the approach used by Muller.\cite{Muller1}

In order to introduce some notation, consider again the standard twin paradox from the point of view of the Earth observer. Suppose the Earth observer and the traveler agree that the traveler's clock begins at the location of the Earth observer's clock with $x'=x=0$ when $t=t'=0$.  Also suppose the Earth twin sees the traveler move to the right until they reach the turning point $x=D$, when the Earth clock reads $T=D/v$. Using the Lorentz transformation we see that if the traveler's clock says this event occurs at a time $T_T$ then
\begin{align}
T_T=\gamma\left(T-\frac{vD}{c^2}\right)&=\gamma\left(\frac{D}{v}-\frac{vD}{c^2}\right)=\frac{D}{v\gamma}\\
\implies T_T&=\frac{T}{\gamma}.
\end{align}
Hence the Earth observer says the traveler's clock will read $T_T=T/\gamma$ and thus that the traveler's clock is running slow by a factor of $\gamma$.  Since the Earth observer knows that acceleration doesn't affect the rate of an ideal clock, they conclude that when the traveler returns to the origin, the Earth clock will read $2D/v$ and the traveler's clock will read $2D/(\gamma v)$.

Putting in some simple numbers will help in what follows. Suppose $v=0.6$ c and $D=3$ c-years (i.e. three light-years).  Then according to the Earth observer the turning point is reached at time
\begin{equation}
T=\frac{D}{v}= \frac{3\; \text{c-years}}{0.6\, \text{c}}=5\;\text{years}.
\end{equation}
The Earth observer predicts that when the traveler reaches the turning point their clock will read 
\begin{equation}\label{travelertime}
T_T=\frac{D}{\gamma v}= \frac{3\; \text{c-year}}{1.25\cdot 0.6\, \text{c}}=4\;\text{year}.
\end{equation}

Consequently, the Earth observer concludes that when the traveling clock returns to the origin and the two clocks are compared, the Earth clock will read
\begin{equation}
\frac{2D}{v}= 10\; \text{year}
\end{equation}
and the traveler's clock will read
\begin{equation}
\frac{2D}{\gamma v}= 8\; \text{year}.
\end{equation}

Now let's consider what happens from the traveler's point of view.  The traveler sees the Earth clock moving to the left until it reaches its turning point at $x_T=-vT_T$.  Using the Lorentz transformation for a frame moving with velocity $-v$, the traveler predicts that when the Earth clock arrives at the turning point it will read
\begin{align}
T=\gamma\left(T_T-\frac{(-v)\cdot x_T}{c^2}\right)&=\gamma\left(T_T-\frac{v^2 T_T}{c^2}\right)\\
\implies T&=\frac{T_T}{\gamma}.\label{1}
\end{align}
Hence the traveler says the clock held by the Earth observer is running slow by a factor of $\gamma$ and we have the makings of a paradox. 

However, let's continue to examine the situation from the traveler's frame. If the distance to the turning point in the Earth frame is $D=3$ c-year then, according to the traveler, that distance is $D_T=D/\gamma= (3$ c-year/$1.25) = 2.4$ c-year.  This means that when the Earth clock reaches the turning point, the traveler's clock reads $T_T= D_T/v= (2.4$ c-year/$0.6$ c) $= 4$ years (as we expected from Eq.({\ref{travelertime}})).  In the traveler's frame the turning point is at $x_T= -D_T$ so, according to the traveler, when the Earth clock reaches this location it will read
\begin{align}
T&=\gamma\left(T_T-\frac{(-v)\cdot (-D_T)}{c^2}\right)\\
&=1.25\left(4\;\text{year} -\frac{(-0.6 \text{c})\cdot (-2.4\;\text{c-year})}{c^2}\right)\label{Eq2}
\\
\implies T&=3.2\; \text{year}\label{Eq2a}
\end{align}
This is just what the traveler expects because in their frame it is the Earth clock that is moving and so must be running slow, as we showed in Eq. (\ref{1}). 

Now the traveler changes into a frame that is returning to the Earth.  In this frame the traveler sees the Earth clock traveling \textit{towards} them with a speed $v=0.6$ c.  We can calculate the reading on the Earth clock measured by the traveler just after the change by using the standard Lorentz transformation.  If the change is quick then the time $T_T$ on the traveler's clock just after the change is essentially the same as the time on the traveler's clock just before the change.  Hence, just after the change, the traveler says the Earth clock reads
\begin{align}
T&=\gamma\left(T_T-\frac{vD_T}{c^2}\right)\label{Eq3},\\
&=1.25\left(4\;\text{year} -\frac{0.6 \text{c}\cdot (-2.4\;\text{c-year})}{c^2}\right)\\
\implies T&=6.8\; \text{year}\label{Eq3a}
\end{align} 
Consequently the traveler measures the Earth clock to read one value just before it changes into an inertial frame traveling back to the Earth and a different value just after.

Now let's consider what the traveler observes \textit{during} this turn-around.  As mentioned in the main body of the paper, if the traveler is moving in a charged spacecraft then the turn-around can occur by having the spacecraft enter a constant magnetic field perpendicular to the plane of its motion.  In this case the speed of the spacecraft is constant as it moves through a semicircle and then heads back towards the Earth.  Both the size of the semicircle and the time it takes for the spacecraft to traverse it can be made small by increasing the strength of the magnetic field.

Modeling the turn-around in this way provides a better understanding of what happens as it occurs.  The general equation for the time transformation is
\begin{equation}\label{gentime}
T=\gamma\left(T_T-\frac{\vct{v}\cdot\vct{x}_T}{c^2}\right),
\end{equation}
where $\vct{v}\cdot\vct{x}=vx\cos{\theta}$.  Because the Earth observer is moving in the $-\uvct{i}$ direction, and because $\vct{x}_T$ points in the $-\uvct{i}$ direction, the traveler observes $\theta=0$ just before the turn-around and $\theta=\pi$ just after.

Since the turn-around is caused by a magnetic field, $\theta=\omega t=(qB/m)t$ and the traveler observers a non-linear speedup of the Earth twin's clock.  The traveler thus observes clocks, and all processes in the Earth frame including the aging of the Earth twin, to speed up during the turn-around.  But did the Earth twin ``really" age more quickly during the turn-around?  In this case there is a ``really" since the Earth twin can be attached to machines that measure the aging process in terms of heart rate, metabolic rate, etc. A digital display of the heartbeat, for example, won't change during the turn-around. In fact, the aging process as viewed from the traveler's frame corresponds to the reading on an accelerometer held by the Earth twin as viewed from the traveler's frame:  just as the accelerometer doesn't register an acceleration, a heart monitor attached to the Earth twin won't register any change.  So, even though the traveling twin observes the Earth twin to accelerate and their clocks to run fast during the turn-around, the Earth twin doesn't ``really accelerate" as measured on an accelerometer and doesn't ``really" age more rapidly as measured on a heart or metabolic monitor.

We can now summarize the complete trip from the traveler's point of view.  Initially the traveler sees the Earth clock moving away from them with a speed $v=0.6$c.  Just as it reaches the turning point, the traveler observes that $4$ years have passed on their clock but only $3.2$ years have passed on the Earth clock.  This is what the traveler expects since the Earth clock is moving in their frame and so is running slow by a factor of $\gamma$.  During the turn-around, the traveler observes the Earth clock to speed up.  At the end of the turn-around, when the Earth clock begins its trip back towards the traveling twin, the traveler says the Earth clock's reading has changed from $3.2$ years to $6.8$ years.  During the return trip, the traveler again observes that $4$ years pass on their clock while only $3.2$ years pass on the Earth clock.  Consequently, when the Earth clock and the traveler's clock meet, the traveler expects the Earth clock to read $6.8 + 3.2$ years $=10$ years!  So the traveler completely understands why the round trip registered $8$ years on their clock and $10$ years on the Earth clock.

What happens during the turn-around can also be understood from the relativity of simultaneity.  As Mermin points out, just before the turn around the traveler observes what the Earth clock reads ``now" from a frame in which the Earth clock is traveling away, while just after the turn around the traveler observes what the Earth clock reads ``now" from a frame in which the Earth clock is returning.\cite{Mermin1}  But two events separated in space that are simultaneous in one frame are not necessarily simultaneous in another. If two events are simultaneous in the Earth frame then, after applying the Lorentz transformation Eq.(\ref{Eq2}), they are not simultaneous in the traveler's frame and the time interval between them is $\gamma vD/c^2$, where $D$ is the separation of the two events in the Earth frame.  The key point is that in Eq.(\ref{Eq2}) the velocity is negative while in Eq.(\ref{Eq3}) it is positive.  Therefore, when we calculate the difference in the reading on the Earth clock before and after the traveler changes frames, we find it is $2\gamma vD/c^2= 2(1.8\, \text{c-year})=3.6\; \text{c-year}$, which is exactly what we found in Eqs.({\ref{Eq2a}}) and ({\ref{Eq3a}}).

Still another way to understand what happens during the turn-around is to note that when viewed from the traveler's frame the clocks in the Earth frame weren't synchronized correctly.  Watching clocks being synchronized in a frame moving away from them, the traveler observes that each successive clock is set out of sync with the clock at the origin by a factor of $\gamma vx/c^2$, where $x$ is the proper distance between the clock being synchronized and the clock at the origin.  When the traveler is in a frame that sees the Earth traveling away in the negative $-\uvct{i}$ direction, they observe that the reading on the Earth clock at the turn-around point is behind what it would have been if the Earth clocks had been synchronized correctly.  (The trailing event -- the reading on the Earth clock at the turn-around point in this case -- occurs first.)  In a frame in which the Earth clock is approaching, the traveler observes that the reading on an Earth clock at the turn-around point is ahead of what it would have been if the Earth clocks had been synchronized correctly.  These two effects add when the traveler changes frames and hence the traveler observes the reading on an Earth clock to advance during the turn-around by $2\gamma vD/c^2= 2(1.8\, \text{c-year})=3.6\; \text{c-year}$.

The same three analyses can be extended to a twin who travels around a polygon.  For example, suppose the Earth observer has set up a system of synchronized clocks in their frame.  When the traveling twin accelerates at each vertex of the polygon they will see the Earth clock at that vertex advance more quickly during the acceleration.  These advances accumulate so that when the traveling twin is again next to the Earth twin both will agree the clock carried around the polygon has recorded less time than the one held by the Earth twin.  Exactly the same thing happens in the limit as the polygon path becomes a circular path.

\section{Appendix III: The General (Standard) form of the Equation for Thomas Precession}

In circular motion with constant speed the acceleration $\vct{a}= (-v^2/r)(\uvct{r})$ and the linear velocity $\vct{v}=\gvct{\omega}\cross\vct{r},$ where $\gvct{\omega}=(2\pi/P)\uvct{k}$ and $P$ is the period.  Consequently
\begin{equation}
\vct{a}\cross\vct{v}=\vct{a}\cross(\gvct{\omega}\cross\vct{r})=\gvct{\omega}(\vct{a}\cdot\vct{r})=-v^2\gvct{\omega}.
\end{equation}
Using this result in Eq.(\ref{TP0}) we see that Thomas Precession of a gyroscope can be expressed in vector form as
\begin{equation}
\gvct{\omega}_T=(\gamma-1)\frac{\vct{v}\cross\vct{a}}{v^2},\label{E5}
\end{equation}
where $\vct{a}$ is the acceleration of the traveling twin and $\vct{v}$ is their velocity. Equation (\ref{E5}) is the exact (and standard) form of the Thomas Precession.  When $v/c$ is small $\gamma\approx 1-v^2/(2c^2)$ and
\begin{equation}
\gvct{\omega}_T=\frac{\vct{a}\cross\vct{v}}{2c^2}+ O\Big(\frac{v^3}{c^4}\Big).\label{E6}
\end{equation}

We can use Eq. (\ref{E6}) to derive the magnitude and direction of the Thomas Precession predicted to occur in GP-B.  Since the satellite is held in orbit by the gravitational force, the acceleration of its reference frame is $\vct{a}=-\big(GM/r^2\big)\uvct{r}$, where $r$ is the radius of the orbit and $M$ is the mass of the Earth.  In this case, since $a=v^2/r,\;\vct{v}=\sqrt{GM/r}\;\uvct{v}$.  Consequently, a gyroscope in the satellite will be observed by the traveling twin to precess with an angular speed
\begin{equation}
\omega_T= \frac{1}{2c^2}\frac{GM}{r^2}\sqrt{\frac{GM}{r}}= \frac{1}{2c^2r}\Big(\frac{GM}{r}\Big)^{3/2}\label{TP1}
\end{equation}
in the direction
\begin{equation}
\guvct{\omega}_T=-\uvct{r}\cross(\uvct{v}),\label{TP2}
\end{equation}
which is anti-parallel to the angular momentum of the satellite in its orbit.  This means that relative to the traveling twin the gyroscope will precess in the opposite sense of the orbit of the satellite.  That is, if the satellite is orbiting the Earth in the counterclockwise direction the traveler will observe the gyroscope to precess in the plane of the orbit in the clockwise direction.

The same analysis can be used understand Thomas Precession in the hydrogen atom.  In this case the ``satellite" is the electron and the ``gyroscope" is the electron's spin vector.  Since the Coulomb force holds the electron in its orbit the acceleration in Eq. (\ref{E5}) is $\vct{a}=-\big(ke^2/m_e r^2\big)\uvct{r}$ where $k$ is a constant, $m_e$ is the mass of the electron and $e$ is the charge of the electron.  Replacing $GM$ in Eq. (\ref{TP1}) with $ke^2/m_e$  the Thomas Precession of the electron's spin vector in its rest frame is
\begin{equation}
{\omega}_T= \frac{1}{2c^2r}\Big(\frac{ke^2}{m_e r}\Big)^{3/2},
\end{equation}
and that the direction of $\gvct{\omega}_T$ is opposite to that of the electron's orbital angular momentum.


\begin{thebibliography}{95}
\bibitem{ftnote3} As is usual in discussions of the twin paradox at the introductory level, we pretend the Earth is an inertial frame and ignore its spin and movement around the Sun.  Although the paradox should be discussed with an observer in an inertial frame rather than on Earth, it loses some of its dramatic appeal when presented in this way.  We will, however, discuss how the Earth's motion is taken into account when we discuss the experiments of Hafele and Keating.
\bibitem{ftnote6} A surface on which the ratio of the circumference of a circle to its radius is less than $2\pi$ has positive curvature, like the surface of a sphere.  The Earth provides a simple example of this situation that is easy to visualize.  Imagine we are looking down on the Earth from the North Pole.  To us, a circle formed by a latitude line above the equator has a radius equal to the length of the longitude line from the North Pole to the circle, which is $R\theta$, where $R$ is the radius of the Earth and $\theta$ is the angle the latitude line makes with the $z$-axis.  On the other hand, the actual radius of the circle is $R\sin\theta$.  Consequently, to an observer looking down from the North Pole, the ratio of the circumference of a latitude line to its radius is $[2\pi R\sin\theta/(R\theta)]< 2\pi$. Note that the geometry is Euclidean when $\theta=\pi/2$, or when the circle is at the equator.
\bibitem{ftnote20} This is the experimental precision quoted on p. 1357 of Newman.\cite{Newman}
\bibitem{ftnote22} This graph is Figure 47 on page 114 of Alley et. al. \cite{Alley}
\bibitem{ftnote5} Just as the famous experiment with muons created in the upper atmosphere is shown in a film \cite{Frisch}, so too are film clips of the experiment of Alley et. al. included in the BBC documentary ``Einstein's Universe," which can be obtained from Corinth Films. \cite{BBC}. Showing both films in class not only provides a welcome change of pace but also reinforces the reality of time dilation.
\bibitem{ftnote9} In class we define a gyroscope as simply a mass spinning around an axis.  We say that the key to the gyroscope is that its mounting can move in any direction without exerting a torque back on the axis about which the mass is spinning.  To the degree that no torques act angular momentum is conserved and the gyroscope points in a fixed direction no matter how its mounting moves.
\bibitem{ftnote11} This derivation is based on a discussion presented in the appendix of Muller, \cite{Muller2} which Muller credits to an idea of Purcell.  The derivation is especially attractive because it only requires knowledge of length contraction whereas most derivations of Thomas Precession are much more complicated (see, e.g., the references in Rhodes and Semon.\cite{Rhodes}
\end{thebibliography}
\end{document}